\documentstyle[sprocl]{article}

\def\be{\begin{equation}}
\def\ee{\end{equation}}
\def\bea{\begin{eqnarray}}
\def\eea{\end{eqnarray}}

\begin{document}

\title{QUANTUM ENTROPY OF CHARGED ROTATING BLACK HOLES}

\author{R.B. MANN}

\address{Dept. of Physics, University of Waterloo, 
Waterloo,\\ ONT N2L 3G1, CANADA}

\maketitle\abstracts{
I discuss a method for obtaining the one-loop 
quantum corrections to the tree-level entropy 
for a charged Kerr black hole.
Divergences which appear can be removed by renormalization of couplings 
in the tree-level gravitational action in a manner similar to that 
for a static black hole.}

\section{Introduction}

For more than two decades physicists have come to appreciate black
holes as thermodynamic systems characterized by
only a few macroscopic parameters such as mass ($m$), charge ($q$)
and angular momentum ($\Omega$). The generic representative of
such a hole in general relativity is the
Kerr-Newman metric \cite{KN}. The thermodynamic analogy 
suggests that there is an entropy associated with 
this hole that is proportional to the area of the event horizon. 

For all other thermodynamic systems, the entropy is proportional
to the logarithm of the number of hidden degrees of freedom.  Are
there analogous degrees of freedom are for a black hole? If so,
where do they come from?
Statistical explanations \cite{4} of their origin
in terms of a gas of quantum fields have been proposed
\cite{6,FS}. Unfortunately the resultant expressions for the entropy
can be understood as one-loop corrections to the classical black hole
entropy, and so do not give any explanation of
the classical entropy itself.

For arbitrary static black holes,
the divergences of the entropy have the same origin as the
UV-divergences of the quantum effective action and can be removed by 
remormalization of the gravitational couplings in the tree-level 
gravitational action\cite{FS}. 
I report here on work carried out with S. Solodukhin
for the analogous problem in the stationary case\cite{RS}.
Remarkably, the UV-divergences for the one-loop entropy
of a Kerr-Newman black hole are renormalized in the same way as 
for a static black hole.

\section{The Euclidean Kerr-Newman Metric}

The Euclidean Kerr-Newman metric can be in the form
\begin{equation}
ds^2_E={\hat{\rho}^2\over \hat{\Delta}}dr^2+{\hat{\Delta}\hat{\rho}^2\over
(r^2-\hat{a}^2)^2}\omega^2+\hat{\rho}^2(d\theta^2+\sin^2\theta \tilde{\omega}^2)
\label{5}
\end{equation}
where the Euclidean time is $t=\imath \tau$ and
the rotation and charge parameters have also been transformed
$a=\imath \hat{a},~q=\imath \hat{q}$, so
that the metric (\ref{5}) is purely real. Here
$\hat{\Delta}(r)=(r-\hat{r}_+)(r-\hat{r}_-)$,
where $r_{\pm}=m\pm \sqrt{m^2+\hat{a}^2+\hat{q}^2}$,
the quantities $\omega$ and $\tilde{\omega}$ take the form 
\begin{equation}
\omega={(r^2-\hat{a}^2)\over \hat{\rho}^2}
(d\tau-\hat{a}\sin^2\theta d \phi ) \qquad
\tilde{\omega}={(r^2-\hat{a}^2)\over 
\hat{\rho}^2}(d\phi+{\hat{a}\over (r^2-\hat{a}^2)}d\tau) 
\label{4}
\end{equation}
with $\hat{\rho}^2=r^2-\hat{a}^2 \cos^2 \theta$.
This space-time has a pair of orthogonal Killing vectors
\begin{equation}
K=\partial_\tau-{\hat{a}\over r^2-\hat{a}^2}\partial_\phi~,
~~\tilde{K}=\hat{a}\sin^2\theta \partial_\tau+\partial_\phi \label{4a}
\end{equation}
which are the respective analogs of the
vectors $\partial_\tau$ and $\partial_\phi$ in the (Euclidean)
Schwarzchild case.
The horizon surface $\Sigma$ defined by $r=\hat{r}_+$ is the stationary 
surface of the Killing vector $K$. Near this surface
the metric (\ref{5}) is approximately
$ds^2_E=ds^2_\Sigma+\hat{\rho}^2_+ ds^2_{C_2}$
where $\hat{\rho}^2_+=\hat{r}^2_+ -\hat{a}^2\cos^2\theta$ and
\begin{equation}
ds^2_\Sigma
=\hat{\rho}_+^2d\theta^2+{(\hat{r}^2_+-\hat{a}^2)^2\over\hat{\rho}_+^2}   
\sin^2\theta d\psi^2
\label{8}
\end{equation}
is the metric on the horizon surface $\Sigma$
up to $O(x^2)$, where $(r-\hat{r}_+)={\gamma x^2 \over 4}$ and
$\gamma=2\sqrt{m^2+\hat{a}^2+\hat{q}^2}$.
 The angle co-ordinate 
$\psi=\phi+{\hat{a} \over (\hat{r}^2_+-\hat{a}^2)} \tau$
and is well-defined on $\Sigma$. 
The metric $ds^2_{C_2}$ is that of a two-dimensional disk $C_2$ 
\begin{equation}
ds^2_{C_2}=dx^2+{\gamma^2 x^2\over 4 \hat{\rho}_+^4}d\chi^2~~.
\label{11}
\end{equation}
attached to $\Sigma$ at a point ($\theta,~\psi$),
where $\chi=\tau-\hat{a}\sin^2\theta\ \phi$ is an angle
co-ordinate on $C_2$.

Regularity of the metric near the horizon implies the identifications
$\psi \leftrightarrow \psi +2\pi$ 
and  $\chi  \leftrightarrow \chi+4\pi\gamma^{-1}\hat{\rho}^2_+$.
For this latter condition to 
hold independently of $\theta$ on the horizon, 
it is also necessary to
identify $(\tau,~\phi )$
with $(\tau+2\pi \beta_H,~\phi-2\pi \Omega \beta_H)$, 
where $\Omega={\hat{a} \over(\hat{r}^2_+- \hat{a}^2)}$ is the
(complex) angular velocity and $\beta_H={(\hat{r}_+^2-\hat{a}^2)/ 
\sqrt{m^2+\hat{a}^2+\hat{q}^2}}$. 
The identified points have the same coordinate $\psi$.

Near $\Sigma$ we therefore have the following description
of the Euclidean  Kerr-Newman geometry:
attached to every point $(\theta, \psi$) 
of the horizon is a two-dimensional disk $C_2$ with coordinates
($x, \chi$). The periodic identification of points on $C_2$ 
holds independently of any point on the horizon $\Sigma$,
even though $\chi$ is not a global coordinate.
As in static case, there is an abelian 
isometry generated by the Killing vector $K$, whose fixed set is
$\Sigma$. Locally we have $K=\partial_\chi$.
The periodicity is in the direction of the vector $K$ and
the resulting Euclidean space $E$ is regular manifold.

Now consider closing the trajectory of $K$ with an arbitrary period
$\beta\neq \beta_H$. This implies the identification
$(\tau+2\pi \beta,~\phi-2\pi \Omega \beta)$, and the metric on
$C_2$ becomes
\begin{equation}
ds^2_{C_{2,\alpha}}=dx^2+\alpha^2x^2d\bar{\chi}^2
\label{12}
\end{equation}
where $\chi=\beta \hat{\rho}^2_+ (\hat{r}_+^2-\hat{a}^2)^{-1}\bar{\chi}$ 
is a new angle coordinate with period $2\pi$. This is
the metric of a two dimensional cone with angular deficit 
$\delta=2\pi (1-\alpha)$, $\alpha\equiv {\beta\over \beta_H}$.
With this new identification the metric (\ref{5}) now describes the 
Euclidean conical space $E_\alpha$ with singular surface $\Sigma$.

For static metrics it is known that curvature tensors behave as
$\alpha$-dependent distribution functions.  This can also be shown
to be true for stationary metrics by regulating (\ref{12}) so
that
$ds^2_{C_{2,\alpha , b}}=f(x,b)dx^2+\alpha^2x^2d\bar{\chi}^2$
where $f(x,b)$ is some smooth regulating function 
such that $\lim_{b\to 0} f(x,b) = 1$.  An evaluation of the
curvature tensors then yields \cite{RS}
\begin{eqnarray}
&&R^{\mu\nu}_{\ \ \alpha\beta} = \bar{R}^{\mu\nu}_{\ \
\alpha\beta}+
2\pi (1-\alpha) \left( (n^\mu n_\alpha)(n^\nu n_\beta)- (n^\mu
n_\beta)
(n^\nu n_\alpha) \right) \delta_\Sigma \nonumber \\
&&R^{\mu}_{ \ \nu} = \bar{R}^{\mu}_{ \ \nu}+2\pi(1-\alpha)(n^\mu
n_\nu)
\delta_\Sigma \quad
R = \bar{R}+4\pi(1-\alpha) \delta_\Sigma
\label{27}
\end{eqnarray}
in the $b \rightarrow 0$ limit,
where $\delta_\Sigma$ is the delta-function
$\int_{\cal M}^{}f\delta_\Sigma = \int_{\Sigma}^{}f$ and 
$(n_\mu n_\nu)=\sum_{a=1}^{2}n^a_\mu n^a_\nu$, where 
$n^a_\mu$ are both normal to $\Sigma$. Barred quantities denoted
tensors evaluated with the unregulated metric (\ref{12}). 
Quadratic curvature invariants may also be shown to
have a structure that is formally identical to the static
case \cite{RS}.

\section{Heat Kernel Expansion}

In the Euclidean path integral approach to a statistical field system 
at temperature $T=(2\pi\beta )^{-1}$ one considers the fields which are 
periodic with
respect to imaginary time $\tau$ with period $2\pi\beta$. 
For a (regulated) rotating black hole 
metric this entails closing the integral curves of $K$ (\ref{4})
for arbitrary $\beta$.  The partition function $Z(\beta)$
then becomes the
functional integral of the matter Euclidean action on $E_\alpha$,
with periodicity conditions imposed on the matter field(s). 

For the matter action 
$I_{E}={1\over 2}\int_{E_\alpha}(\nabla \varphi )^2$
standard techniques yield
\begin{equation}
\ln Z (\beta )=-{1\over 2}\ln det (-\Box_{E_\alpha})
{1\over (4\pi s)^2}\sum^{\infty}_{n=0} a_n s^n
\label{40}
\end{equation}
where the heat-kernel coefficients $a_n$
are a sum of standard and conical coefficients. An evaluation
of these indicates that the UV-divergent part of the entropy is
renormalized in a manner identical to that for static black holes.
Once this is taken into account, the contribution to the entropy
$S=-(\beta \partial_\beta-1)\ln Z(\beta )|_{\beta=\beta_H}$
for a Kerr-Newman black hole becomes \cite{RS}
\begin{equation}
S_{div}={1 \over 48\pi \epsilon^2}  A_\Sigma 
+{1\over 45}
\left( 1 -{3q^2\over 4r^2_+}(1+{(r^2_++a^2)\over a r_+} \tan^{-1}({a\over r_+}))
\right) \ln {L \over \epsilon}
\label{46}
\end{equation}
in the Lorentzian section,
where $A_\Sigma=4\pi (r^2_++a^2)$ is area of the horizon $\Sigma$.
When $q=0$ the quantum-corrected part of the
entropy in (\ref{46}) is the same as that for
a Schwarzchild black hole. 
At present there is no explanation for this.

\section{Concluding Remarks}

It is still an open question as to
what degrees of freedom are counted by the entropy 
of black hole. The conical-deficit methods employed here clearly indicate
that the entropy of a Kerr-Newman black hole is associated with
the horizon. A proper treatment of the statistical-mechanical
calculation of the quantum entropy 
should provide us with a better understanding of the 
relationship between the different methods of assigning entropy 
to a black hole.

\section*{Acknowledgements}
I would like to thank the organizers of the 2nd Sakharov Conference
for their invitation to me to speak at this meeting and for their
kind hospitality.
This work was supported by the Natural Sciences and Engineering Research
Council of Canada.

\end{document}